\pdfoutput=1
\documentclass[aps,prl,twocolumn,amsmath,amssymb,showpacs,superscriptaddress,notitlepage,longbibliography,floatfix]{revtex4-1}

\usepackage[colorlinks=true,linkcolor=blue,anchorcolor=red,citecolor=blue, urlcolor=blue]{hyperref}
\usepackage{bm}
\usepackage{graphicx}
\usepackage{xcolor}
\usepackage{cases}

\begin{document}

\title{Direct imaging of vortex lattices across fractional fillings in two-dimensional Josephson junction arrays}

\author{Cequn Li}
\affiliation{Laboratory of Atomic and Solid-State Physics, Cornell University, Ithaca, NY 14853, USA}

\author{Austin R. Kaczmarek}
\affiliation{Laboratory of Atomic and Solid-State Physics, Cornell University, Ithaca, NY 14853, USA}

\author{Omri Lesser}
\affiliation{Laboratory of Atomic and Solid-State Physics, Cornell University, Ithaca, NY 14853, USA}

\author{Maciej W. Olszewski}
\affiliation{Department of Physics, Cornell University, Ithaca, NY 14853, USA}

\author{Katja C. Nowack}
\email{kcn34@cornell.edu}
\affiliation{Laboratory of Atomic and Solid-State Physics, Cornell University, Ithaca, NY 14853, USA}
\affiliation{Kavli Institute at Cornell for Nanoscale Science, Ithaca, NY 14853, USA}

\begin{abstract}
Josephson junction arrays are predicted to host a rich sequence of commensurate vortex lattices. However, direct imaging of these configurations remains limited, while transport measurements cannot resolve the underlying vortex arrangements or how order breaks down. Using scanning superconducting quantum interference device (SQUID) microscopy, we resolve individual vortices and directly image this long-predicted hierarchy spanning filling factors $f=1/8$ to $1/2$ in two-dimensional Au-Nb Josephson junction arrays, including several states with no discernible transport signatures. At $f=1/4$, two structurally distinct vortex orders with $4\times4$ and $2\times4$ unit cells coexist, consistent with their energetic degeneracy in the frustrated XY model. Away from commensurate filling, we resolve defect structures tied to the underlying vortex-lattice order: the $f=1/2$ checkerboard retains its order around isolated excess vortices and vacancies, whereas the $f=1/3$ staircase fragments into domains separated by domain walls. These results reveal how frustration, competing configurations, and defects shape the spatial coherence of commensurate vortex order in engineered superconducting lattices.
\end{abstract}

\maketitle
The formation of vortex lattices illustrates how repulsive interactions can produce collective order. In artificial lattices such as two-dimensional (2D) Josephson junction arrays (JJAs) \cite{resnick1981kosterlitz,teitel1983josephson,halsey1985josephson,rzchowski1990vortex,falo1990v,kolahchi1991ground,straley1993phase,theron1994observation,van1996quantum,eley2012approaching,han2014collapse,poccia2015critical,cosmic2018circuit,bottcher2018superconducting,martinoli2000two,korshunov2001vortex,frolov2008imaging,tesei2009magnetic,lankhorst2018scaling,naibert2021imaging,bishop2022local,penner2023resistivity,bottcher2023dynamical,lesser2024josephson,reinhardt2024magnetochiral,vervoort2025dc,sasmal2025voltage,teller2025frustrated,Mikalsen2026,wang2026long,Bondar2026,BishopVanHorn2026} and superconducting wire networks \cite{hallen1993direct,vu1993imaging,ling1996nature,abilio1999magnetic,wang2024flux}, vortex-vortex interactions compete with the periodic lattice potential to generate commensurate vortex configurations \cite{teitel1983josephson,halsey1985josephson,hallen1993direct,vu1993imaging,wang2024flux}. 
In 2D JJAs, an out-of-plane magnetic field $B$ introduces frustration characterized by the filling factor $f=BA/\Phi_0$, where $A$ is the plaquette area and $\Phi_0=\frac{h}{2e}$ is the superconducting flux quantum, with $h$ the Planck constant and $e$ the electron charge. Such arrays can be mapped onto the frustrated XY model \cite{Kosterlitz1973,resnick1981kosterlitz,teitel1983josephson,halsey1985josephson}. At rational $f$, theory predicts a hierarchy of commensurate vortex lattices whose structures depend sensitively on filling and lattice geometry, including lattices with large unit cells and staircase-like states that approach the checkerboard configuration at $f=1/2$ \cite{teitel1983josephson,halsey1985josephson,kolahchi1991ground,straley1993phase}. The lattice geometry and junction coupling can be tuned lithographically or electrostatically \cite{han2014collapse,bottcher2018superconducting,sasmal2025voltage,wang2026long,Mikalsen2026,wang2026long,Bondar2026,BishopVanHorn2026}. This control makes JJAs a platform for engineering frustrated vortex matter and exploring collective many-body phenomena \cite{poccia2015critical,lankhorst2018scaling,bottcher2023dynamical,han2014collapse,bottcher2018superconducting,sasmal2025voltage}, with proposals to realize topological superconductivity \cite{lesser2024josephson,ioffe2002possible}.

Experimentally, commensurate vortex states have been inferred primarily through global signatures at magnetic fields corresponding to rational fillings, such as enhancements of the critical current~\cite{poccia2015critical,bottcher2023dynamical,reinhardt2024magnetochiral,vervoort2025dc,sasmal2025voltage,teller2025frustrated} or diamagnetic response~\cite{theron1994observation,tesei2009magnetic}. These signatures indicate commensurability but do not directly reveal the vortex arrangement, competing local orders, or the defects and domain walls (DWs) through which order is lost. Although real-space imaging of vortices in JJAs and superconducting networks has been demonstrated previously \cite{frolov2008imaging,hallen1993direct,vu1993imaging,wang2024flux,wang2026long}, most studies have focused on the robust $f=1/3$ and $1/2$ states, which exhibit strong transport signatures. Direct imaging of the broader sequence of predicted vortex lattices remains limited, particularly for complex configurations with large unit cells and those that produce weak or absent signatures in global measurements.

Here, we use scanning SQUID microscopy to directly image commensurate vortex configurations in 2D square Au-Nb JJAs, spanning fillings from $f=1/8$ to $1/2$. We resolve complex configurations at $f=3/10$, $3/8$, and $3/7$, including states with no discernible signatures in transport. At $f=1/4$, the images reveal coexistence of distinct $4\times4$ and $2\times4$ vortex orders that are energetically degenerate in the frustrated XY model. For small deviations from  commensurate filling, we observe distinct ways in which defects disrupt long-range order in the $f=1/2$ and $1/3$ states. These measurements provide direct microscopic insight into how competing configurations and defect formation shape commensurate vortex order in frustrated artificial lattices.

\begin{figure*}[tbph]
\centering \includegraphics[width=0.9\textwidth]{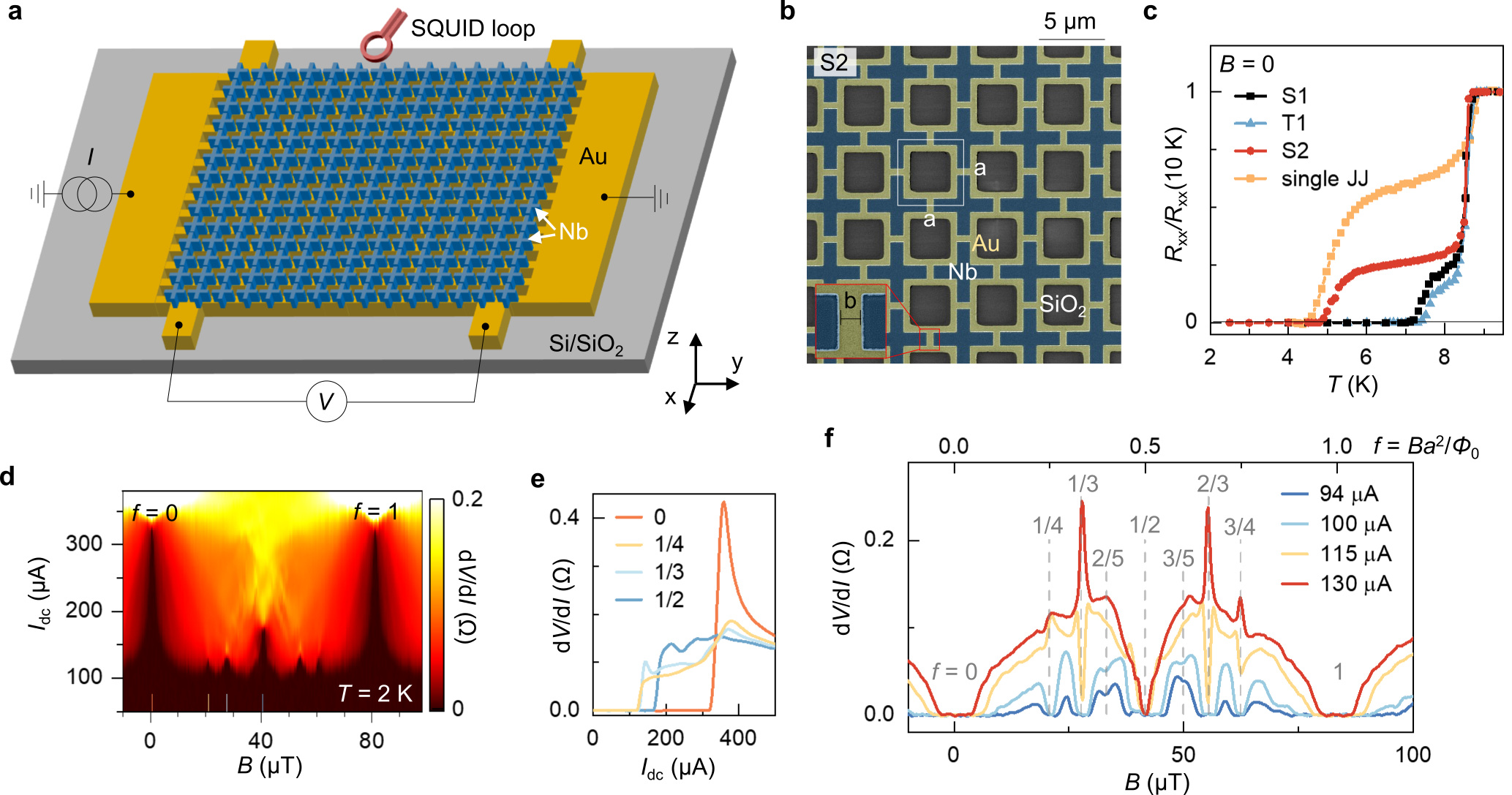}
\caption{\textbf{Experimental setup and transport signatures of commensurate vortex states in Au-Nb JJAs.} (a) Schematic of the device and measurement setup. The square arrays studied here contain $N_x=25$ by $N_y=30$ plaquettes. The SQUID sensor, with pickup-loop inner and outer radii of 0.75 and $1.6~\mu\mathrm{m}$, respectively, is scanned approximately 1 \textmu m above the array. Additional details on the scanning SQUID setup are provided in Supplemental Material~\cite{SupplementalMaterial}, Sec. S1. (b) False-colored SEM image of device S2. The dashed white box marks one plaquette. The inset shows a magnified view of a single Josephson junction with $b=400~\mathrm{nm}$. (c) Normalized longitudinal resistance $R_{xx}$ as a function of temperature for multiple devices. Devices T1 and S1 have $b=150~\mathrm{nm}$, with T1 having a triangular lattice geometry. A single junction with the same $b$ as device S2 is shown for comparison. Despite differences in lattice constant and geometry, devices with the same $b$ exhibit similar $T_c^\mathrm{arr}$. (d) Differential resistance $dV/dI$ as a function of out-of-plane magnetic field $B$ and dc bias current $I_{\mathrm{dc}}$. A larger-scale map is given in Supplemental Material~\cite{SupplementalMaterial}, Fig. S3, revealing additional oscillation periods at higher integer and half-integer fillings. (e) $\mathrm dV/\mathrm dI$ versus $I_{\mathrm{dc}}$ at selected fillings indicated by the vertical lines in (d). (f) $\mathrm dV/\mathrm dI$ versus \(B\) at selected values of $I_{\mathrm{dc}}$. Data in (d)–(f) are from device S2 at $T=2~\mathrm{K}$.}
\label{fig:setup}
\end{figure*}

Figure~\ref{fig:setup}(a) illustrates the device structure of our 2D Au-Nb JJAs. We fabricate the devices by patterning a Ti/Au (5/40 nm) Hall bar on a Si/SiO$_2$ substrate, followed by sputtering Ti/Nb (3/70 nm) islands \cite{Olszewski2026} onto the Ti/Au layer (see Supplemental Material~\cite{SupplementalMaterial}, Sec.~S2 for details). Figure~\ref{fig:setup}(b) shows a false-colored scanning electron microscopy (SEM) image of a representative square JJA, device S2. The two square arrays studied here, S1 and S2, have lattice constants $a=4.75$ and $5~\mu\mathrm{m}$, and inter-island separations $b=150$ and $400~\mathrm{nm}$, respectively.

We first characterize the JJAs by measuring the temperature dependence of the longitudinal resistance, $R_{xx}(T)$, at zero magnetic field. As shown in Fig.~\ref{fig:setup}(c), $R_{xx}$ drops at the Nb superconducting transition temperature, $T_c^\mathrm{Nb}=8.7$ K, followed by a lower-temperature shoulder associated with the onset of phase coherence across the array \cite{eley2012approaching}. The zero-resistance array transition temperature $T_c^\mathrm{arr}$ depends primarily on the inter-island separation $b$, ranging from $\sim7.1$ K for $b=150~\mathrm{nm}$ to $\sim$ 4.7 K for $b=400~\mathrm{nm}$.


\begin{figure*}[tbph]
\centering \includegraphics[width=0.9\textwidth]{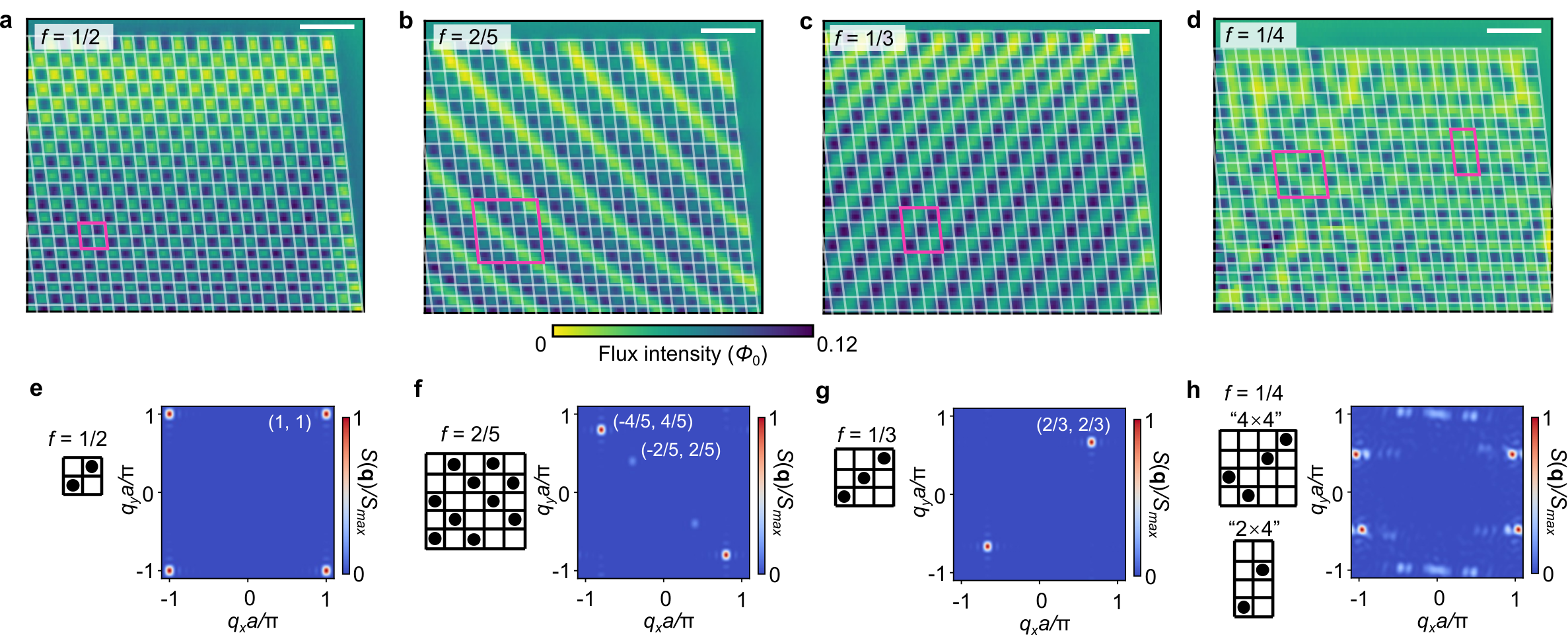}
\caption{\textbf{Vortex configurations associated with transport signatures.} Flux images acquired at (a) $f=1/2$, (b) $2/5$, (c) $1/3$, and (d) $1/4$ at $T=0.2$~K. The flux intensities are offset to reflect the negligible flux due to Meissner screening in the bright-yellow region. White solid lines indicate the array lattice for clarity. The magenta boxes mark representative ordered regions used to identify the vortex-lattice unit cells. The images shown here were acquired in device S2 and are reproducible in device S1. Scale bars, 20 \textmu m. (e)--(h) Corresponding vortex-lattice unit cells and structure factors $S(\textbf{q})$ within the first Brillouin zone for the configurations in (a)--(d), respectively.}
\label{fig:vortex_lattice_1}
\end{figure*}

\begin{figure*}[tbph]
\centering \includegraphics[width=0.9\textwidth]{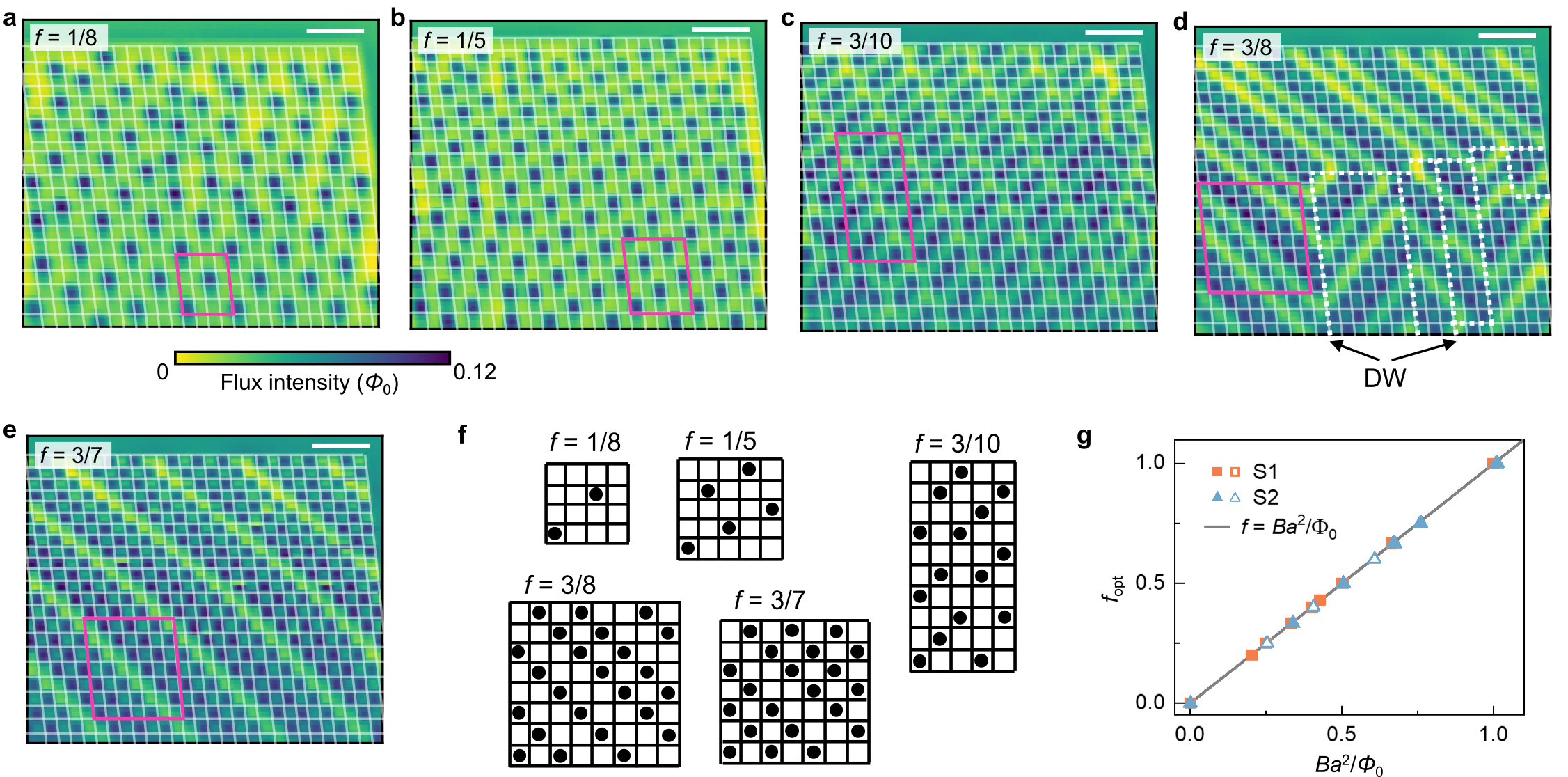}
\caption{\textbf{Commensurate vortex states without distinct transport signatures.} Flux images acquired near (a) $f=1/8$, (b) $1/5$, (c) $3/10$, (d) $3/8$, and (e) $3/7$ at $T=0.2$~K. White dashed lines mark DWs separating locally ordered regions,  as illustrated for $f=3/8$. Images are from device S1. Scale bars, 20 \textmu m. (f) Schematics of the corresponding vortex-lattice unit cells extracted from (a)--(e). (g) Commensurate fillings $f_\mathrm{opt}$ identified from real-space imaging in (a)--(e) and Figs.~\ref{fig:vortex_lattice_1}(a)--(d) (solid symbols), and from transport measurements in Fig.~\ref{fig:setup}(f) and Supplemental Material~\cite{SupplementalMaterial}, Fig. S3(b) (open symbols), plotted as a function of $Ba^2/\Phi_0$. The gray solid line indicates the expected relation $f=Ba^2/\Phi_0$.}
\label{fig:vortex_lattice_2}
\end{figure*}

Figures~\ref{fig:setup}(d)--(f) summarize the transport signatures of commensurate vortex states in device S2. The critical current $I_c$ oscillates periodically with magnetic field $B$, with a period $B_0\approx83~\mu\mathrm{T}$, in good agreement with $B_0=\Phi_0/a^2=82.7~\mu\mathrm{T}$ for $a=5~\mu\mathrm{m}$. Strong commensurability features appear at integer and half-integer fillings, with additional features at $f=1/4$, $1/3$, $2/5$, $2/3$, and $3/4$. With increasing $I_{\mathrm{dc}}$, the differential resistance minima at $f=1/4$, $1/3$, and $2/5$ evolve toward maxima, consistent with current-driven depinning or melting of commensurate vortex states \cite{poccia2015critical,lankhorst2018scaling,bottcher2023dynamical}. Similar results are obtained in device S1 (see Supplemental Material~\cite{SupplementalMaterial}, Fig. S3). Several additional ordered states identified by real-space imaging below, however, show no distinct transport signatures, highlighting the limited sensitivity of global transport to weaker vortex-lattice order.

\begin{figure*}[tbph]
\centering \includegraphics[width=0.9\textwidth]{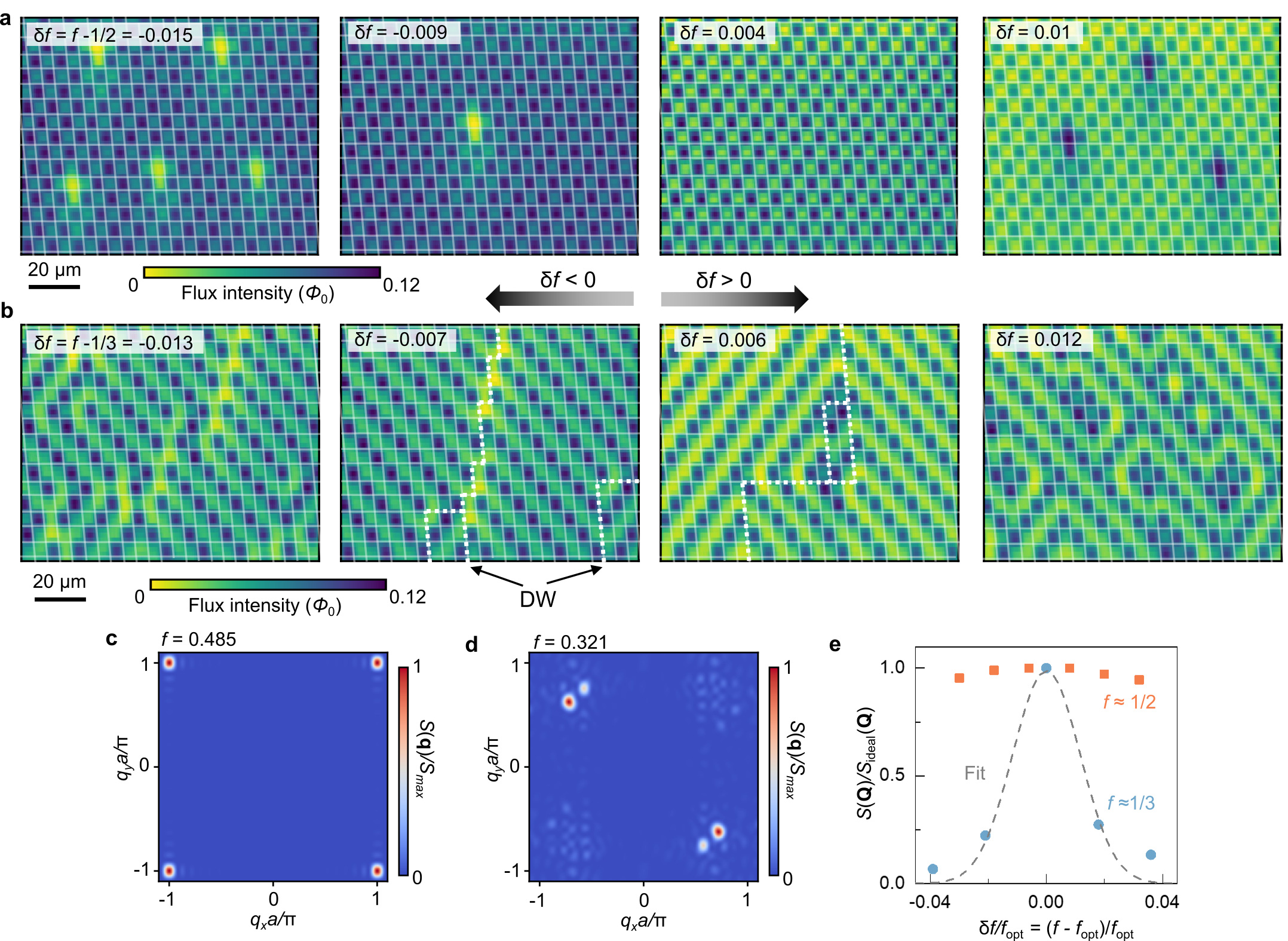}
\caption{\textbf{Evolution of vortex-lattice order under small deviations from commensurate filling.} (a) Flux images acquired near $f=1/2$, with $\delta f=f-1/2$. (b) Flux images acquired near $f=1/3$, with $\delta f=f-1/3$. White dashed lines mark DWs separating locally ordered regions. Images are from device S1. $T=0.2$~K. (c) and (d) Structure-factor maps $S(\mathbf q)$ corresponding to the representative states at $f=0.485$ and $0.321$, respectively. (e) Normalized principal Bragg-peak intensity, $S(\mathbf Q)/S_{\mathrm{ideal}}(\mathbf Q)$ as a function of $\delta f/f_\mathrm{opt}$. The gray dashed curve is an empirical Gaussian fit to the $f\approx1/3$ data, $\exp\left[-(\delta f/f_\mathrm{opt})^2/(2\sigma^2)\right]$, with $\sigma=0.012$.}
\label{fig:DW}
\end{figure*}

To visualize the vortex configurations, we perform scanning SQUID microscopy in the absence of a transport current through the array. Because vortex mobility and magnetic screening vary strongly with temperature \cite{tesei2009magnetic}, the array is field-cooled to $T<1~\mathrm{K}$ at a rate of $\sim1$ K/min at each magnetic field before imaging, promoting relaxation toward low-energy configurations during cooling while minimizing sensor-induced rearrangement during imaging. The vortex trapping effect in the Nb islands is negligible because of their narrow trace width and the small magnetic-field range used in our measurements \cite{Stanprl2004,Baiprb2026}. Detailed discussions on vortex mobility, temperature-dependent magnetic contrast, and imaging protocol are given in Supplemental Material~\cite{SupplementalMaterial}, Secs.~ S4 and S5 \cite{BishopVanHorn2022}.

Guided by the transport measurements, we acquired image series across $f=1/4$, $1/3$, $2/5$, and $1/2$ in device S2, and show representative vortex configurations in Fig.~\ref{fig:vortex_lattice_1}. In each image, the bright yellow regions indicate reduced flux due to Meissner screening, whereas the dark spots correspond to vortices. At $f=1/3$, $2/5$, and $1/2$, we frequently observe nearly single-domain vortex lattices spanning the entire field of view with symmetries consistent with the predicted lowest-energy states \cite{teitel1983josephson}. In contrast, the vortex configuration at $f=1/4$ is less ordered and contains two competing local motifs, referred to as the $4\times4$ and $2\times4$ configurations, as illustrated in Fig.~\ref{fig:vortex_lattice_1}(h). Unlike the symmetry-related degenerate configurations observed at other fillings, these two configurations are structurally distinct. Quantitative analysis of multiple images yields average ordered-area fractions of  $33\pm4\%$ and $26\pm3\%$ for the $2\times4$ and $4\times4$ configurations, respectively, with the remaining $41\pm2\%$ classified as disordered or ambiguous (see Supplemental Material~\cite{SupplementalMaterial}, Sec.~S8). 

We perform zero-temperature Monte Carlo simulations of the uniformly frustrated XY model. We find that both the $2\times 4$ and $4\times4$ configurations have an energy per site of $-\frac{1}{2}(1+\sqrt{3})E_J$, where $E_J$ is the Josephson coupling energy of a single junction (see Supplemental Material~\cite{SupplementalMaterial}, Sec.~S9). This degeneracy is consistent with prior theoretical work~\cite{straley1993phase} and with the experimentally observed coexistence of the two orders. Their comparable ordered-area fractions are consistent with neither configuration being strongly favored under our experimental conditions. Additional calculations show that weak directional anisotropy in the Josephson couplings can lift this degeneracy in favor of the $2\times4$ configuration, which could explain its slightly larger observed area fraction. Competition between the two configurations produces substantial disorder near their domain boundaries.

The symmetry and spatial coherence of these vortex configurations are also reflected in reciprocal space. We calculate the static structure factor \(S(\mathbf q)\) from the binary vortex-occupancy maps by Fourier transforming the density fluctuations \cite{DennistonTang1998} (see details in Supplemental Material~\cite{SupplementalMaterial}, Sec.~S7). Periodic vortex order produces Bragg peaks in reciprocal space, whose positions identify the underlying lattice symmetry and whose intensities quantify the degree of long-range order. As shown in Figs.~\ref{fig:vortex_lattice_1}(e)--(g), the nearly single-domain configurations at $f=1/2$, $2/5$, and $1/3$ exhibit pronounced Bragg peaks at the principal ordering wavevectors expected for the corresponding ideal vortex lattices. For the $f=1/3$ and $2/5$ states, we also observe symmetry-related, energetically degenerate configurations in which the principal stripe direction is rotated by $90^\circ$, resulting in a corresponding $90^\circ$ ($C_4$) rotation of the Bragg-peak pattern in reciprocal space (see Supplemental Material~\cite{SupplementalMaterial}, Fig. S7). 

The coexistence of the two orders at $f=1/4$ is also apparent in reciprocal space, as shown in Fig.~\ref{fig:vortex_lattice_1}(h). Its structure factor exhibits two sets of Bragg peaks near the characteristic ordering wavevectors of the $4\times4$ and $2\times4$ configurations (see Supplemental Material~\cite{SupplementalMaterial}, Fig.~S8). The peaks are substantially broader
than those of the nearly single-domain states and are accompanied by additional diffuse intensity, indicating reduced spatial coherence and enhanced disorder associated with the competing configurations and their domain boundaries.

Imaging resolves additional vortex configurations at $f=1/8$, $1/5$, $3/10$, $3/8$, and $3/7$ that are not detected in transport as shown in Figs.~\ref{fig:vortex_lattice_2}(a)--(e), respectively. These states are more fragile and typically form multidomain structures with DWs, indicating reduced long-range coherence consistent with their weak or absent transport signatures~\cite{wang2026long}. For example, Fig.~\ref{fig:vortex_lattice_2}(e) shows one of the most ordered $f=3/7$ configurations observed, whereas repeated thermal cycles more commonly yield imperfect patterns (see Supplemental Material~\cite{SupplementalMaterial}, Fig.~S6). At $f=3/8$, we observe domains of energetically degenerate, $C_4$-symmetry-related configurations, separated by DWs (Fig.~\ref{fig:vortex_lattice_2}(d)). The structure factors of these states nevertheless retain characteristic Bragg-peak patterns that identify the underlying vortex-lattice symmetry and agree with those calculated for the corresponding ideal configurations (see Supplemental Material~\cite{SupplementalMaterial}, Fig. S9). Figure~\ref{fig:vortex_lattice_2}(f) summarizes the corresponding vortex-lattice unit cells extracted from the real-space images, in agreement with theoretical predictions: $f=1/q$ states with integer $q\geq5$ form simple, regularly spaced vortex lattices, whereas the states in $1/3\leq f<1/2$ exhibit more complex staircase-like structures \cite{teitel1983josephson,kolahchi1991ground}. Figure~\ref{fig:vortex_lattice_2}(g) shows that the fillings identified from the real-space patterns follow the expected relation $f=Ba^2/\Phi_0$.

The real-space images further reveal how commensurate order breaks down away from optimal filling $f_\mathrm{opt}$. We focus on the robust $f=1/2$ and $1/3$ states and quantify long-range order using the normalized principal Bragg-peak intensity, $S(\mathbf Q)/S_{\mathrm{ideal}}(\mathbf Q)$, where $\mathbf Q$ is the ordering wavevector and $S_{\mathrm{ideal}}(\mathbf Q)$ is calculated for an ideal, defect-free configuration over the same field of view (see details in Supplemental Material~\cite{SupplementalMaterial}, Sec.~S7). Near $f=1/2$, excess vortices or vacancies appear as localized defects, while the surrounding checkerboard order remains largely intact (Fig.~\ref{fig:DW}(a)), as evidenced by the preserved Bragg-peak pattern (see an example in Fig. ~\ref{fig:DW}(c)) and the gradual decrease of the peak intensity at $\mathbf Q=(\pi/a,\pi/a)$ (Fig. ~\ref{fig:DW}(e)). Only at larger deviations is the long-range order substantially suppressed, consistent with the weakening of the $f=1/2$ commensurability feature in transport. By contrast, even small deviations from $f=1/3$ generate DWs that partition the lattice into locally ordered regions (Fig.~\ref{fig:DW}(b)). The vortex arrangements across these DWs resemble those of the low-energy herringbone-type excitations previously identified theoretically at $f=1/3$ \cite{DennistonTang1998}. Accompanying this domain formation, the Bragg-peak intensity drops rapidly, with an empirical Gaussian fit yielding a width $\sigma\approx0.012$ in the normalized filling deviation $\delta f/f_\mathrm{opt}$ (Fig.~\ref{fig:DW}(e)).

The distinct defect structures likely reflect the different symmetry and degeneracy of the parent commensurate states: the $f=1/2$ checkerboard accommodates localized defects without immediately disrupting the surrounding order, whereas the $f=1/3$ state consists of diagonal staircases with symmetry-related orientations, allowing extended DWs to separate locally ordered staircase domains. More generally, the more fragile states at $f=1/8$, $1/5$, $3/10$, $3/8$, and $3/7$ possess larger or more complex unit cells and, in several cases, multiple symmetry-related realizations. These additional ordering degrees of freedom may provide natural channels for multidomain formation and DWs, consistent with the broadened Bragg peaks and reduced long-range coherence observed near their optimal fillings. Their reduced coherence likely contributes to the weak or absent transport signatures~\cite{wang2026long}.

In summary, we directly image a broad sequence of long-predicted vortex lattices and reveal competing local orders and defects that global transport cannot resolve. The coexistence of distinct degenerate orders at $f=1/4$, together with domain formation among symmetry-related states and the contrasting defect structures near $f=1/2$ and $1/3$, highlights how lattice symmetry, competition among multiple ordered configurations, and domain-wall formation govern the stability and breakdown of commensurate vortex order.

\begin{acknowledgements} {\emph{Acknowledgments}}-- C.L., A.R.K., and K.C.N. acknowledge support from the Air Force Research Laboratory, Project Grant FA9550-21-1-0429. O.L. acknowledges support from the Bethe-KIC postdoctoral fellowship at Cornell University. Nb film deposition development was supported by the Air Force Office of Scientific Research under award number FA9550-23-1-0706. Any opinions, findings, and conclusions or recommendations expressed in this material are those of the author(s) and do not necessarily reflect the views of the United States Air Force. The device fabrication was performed in part at the Cornell NanoScale Facility, an NNCI member supported by NSF Grant NNCI-2025233. We thank Daniel C. Ralph and Valla Fatemi for sharing the sputter system for Nb film deposition. We thank Joris Van de Vondel, Simon Reinhardt, and Debanjan Chowdhury for helpful discussions.
\end{acknowledgements}

\noindent\textit{Author contributions}-- K.C.N. conceived the experiment; C.L. fabricated the devices and performed the transport measurements; C.L. and  A.R.K. performed the scanning SQUID measurements under the supervision of K.C.N.; C.L. performed the structure factor analysis, and A.R.K. performed supercurrent simulations under the supervision of K.C.N.; O.L. performed the Monte Carlo simulations; M.W.O. sputtered Nb; C.L. and K.C.N. wrote the manuscript with input from all authors.

\noindent\textit{Data availability}-- The data that support the findings of this study are available from the corresponding author upon reasonable request.

\bibliographystyle{apsrev4-1-etal-title}
\bibliography{refs-JJAs.bib}

\clearpage
\onecolumngrid

\begin{center}
{\large\bfseries  Supplemental Material:\\[0.25em]
Direct imaging of vortex lattices across fractional fillings in two-dimensional Josephson junction arrays\par}

\vspace{1.0em}

{Cequn Li$^{1}$, Austin R. Kaczmarek$^{1}$, Omri Lesser$^{1}$, Maciej W. Olszewski$^{2}$, and Katja C. Nowack$^{1,3,*}$\par}

\vspace{0.75em}

{\small
$^{1}$Laboratory of Atomic and Solid-State Physics, Cornell University, Ithaca, NY 14853, USA\\
$^{2}$Department of Physics, Cornell University, Ithaca, NY 14853, USA\\
$^{3}$Kavli Institute at Cornell for Nanoscale Science, Ithaca, NY 14853, USA\\[0.5em]
$^{*}$\href{mailto:kcn34@cornell.edu}{kcn34@cornell.edu}\par}
\end{center}

\vspace{0.5em}

\makeatletter
\newcommand{\sitableofcontents}{\section*{Contents}\@starttoc{sitoc}}
\newcommand{\sisection}[1]{\section{#1}\addcontentsline{sitoc}{section}{\protect\numberline{\thesection}#1}}
\newcommand{\sisubsection}[1]{\subsection{#1}\addcontentsline{sitoc}{subsection}{\protect\numberline{\thesubsection}#1}}
\makeatother

\setcounter{section}{0}
\setcounter{subsection}{0}
\setcounter{secnumdepth}{2}
\setcounter{equation}{0}
\setcounter{figure}{0}
\setcounter{table}{0}
\setcounter{page}{1}
\renewcommand{\thepage}{S\arabic{page}}
\renewcommand{\thesection}{S\arabic{section}}
\renewcommand{\theequation}{S\arabic{equation}}
\renewcommand{\thefigure}{S\arabic{figure}}

\makeatletter
\renewcommand*{\theHsection}{SI.\arabic{section}}
\renewcommand*{\theHsubsection}{SI.\arabic{section}.\arabic{subsection}}
\renewcommand*{\theHequation}{SI.\arabic{equation}}
\renewcommand*{\theHfigure}{SI.\arabic{figure}}
\renewcommand*{\theHtable}{SI.\arabic{table}}
\makeatother

\sitableofcontents

\begin{figure*}[tbph]
\centering \includegraphics[width=0.7\textwidth]{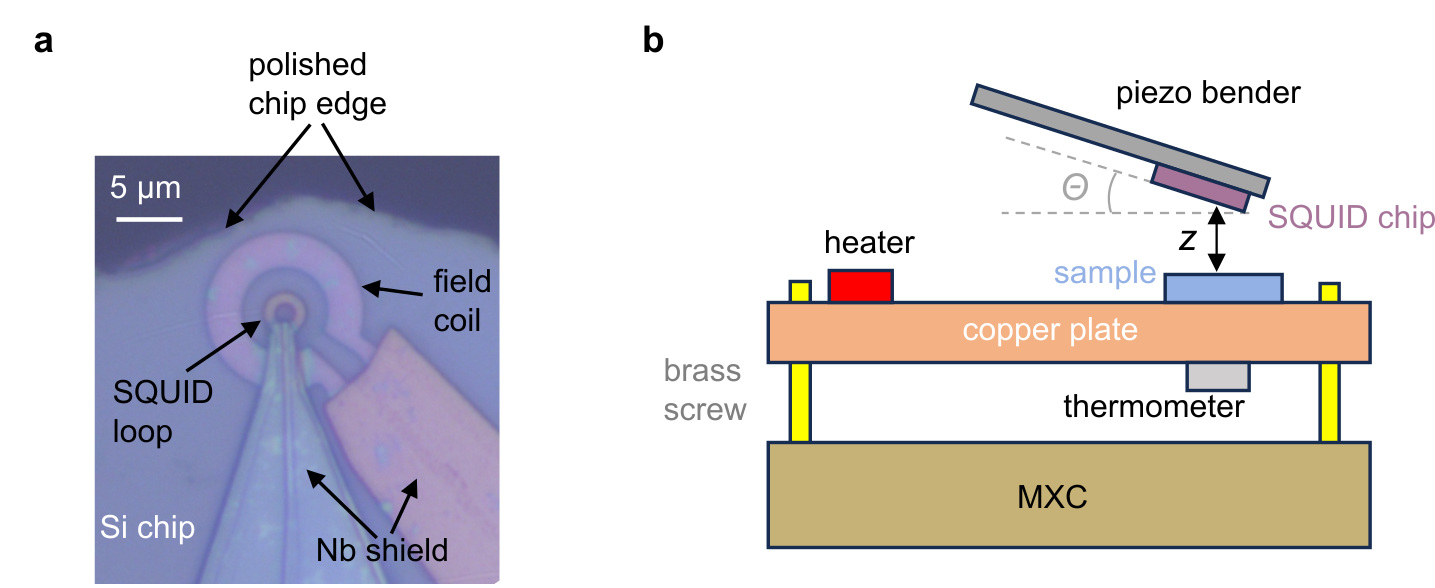}
\caption{\textbf{Experimental setup.} (a) Optical image of the scanning SQUID magnetometer used in this work. (b) Schematic of the setup used for scanning SQUID magnetometry measurements (not to scale). The SQUID chip was mounted on a piezo bender and approached the sample at a small angle $\theta\approx5^\circ$.}
\label{fig:squid}
\end{figure*}

\clearpage
\sisection{Scanning SQUID setup}
Figure~\ref{fig:squid}(a) shows an optical image of the SQUID magnetometer used in this work. The sensor employs the same gradiometric design described in Ref.~\cite{Huber2008}, with a pickup loop having inner and outer radii of 0.75 and $1.6~\mu\mathrm{m}$, respectively. The SQUID is coupled to a SQUID-array amplifier mounted on the mixing chamber (MXC) of the dilution refrigerator for low-noise readout; further details of the readout scheme are given in Ref.~\cite{Huber2008}. The field coil was not used in this study.

Figure~\ref{fig:squid}(b) schematically illustrates our experimental setup. The Josephson junction array (JJA) chips were mounted on a copper sample plate thermally anchored to the MXC. We used thin brass screws as thermal links between the sample plate and the MXC to vary the sample temperature from the Nb transition temperature, $T_c$ $\sim$ 9 K, to $T=200$ mK without substantially heating the MXC. A Cernox thermometer was mounted beneath the sample, while a heater was positioned away from both the sample and thermometer to reduce thermal gradients and lag. The SQUID magnetometer was mounted on a separate stage controlled by piezoelectric positioners and scanners, with a typical SQUID-sample separation of $\sim1~\mu\mathrm{m}$ during imaging. The cryostat was magnetically shielded by a mu-metal can. The out-of-plane background magnetic field was tuned using a Helmholtz coil mounted inside the cryostat, with a field constant of $\sim282~\mathrm{\mu T/A}$, providing the fine field control required to resolve commensurate vortex configurations. The sample stage also provided dc electrical connections for transport measurements, with the JJA contact pads wire-bonded to the measurement leads.

\begin{figure*}[tbph]
\centering \includegraphics[width=0.9\textwidth]{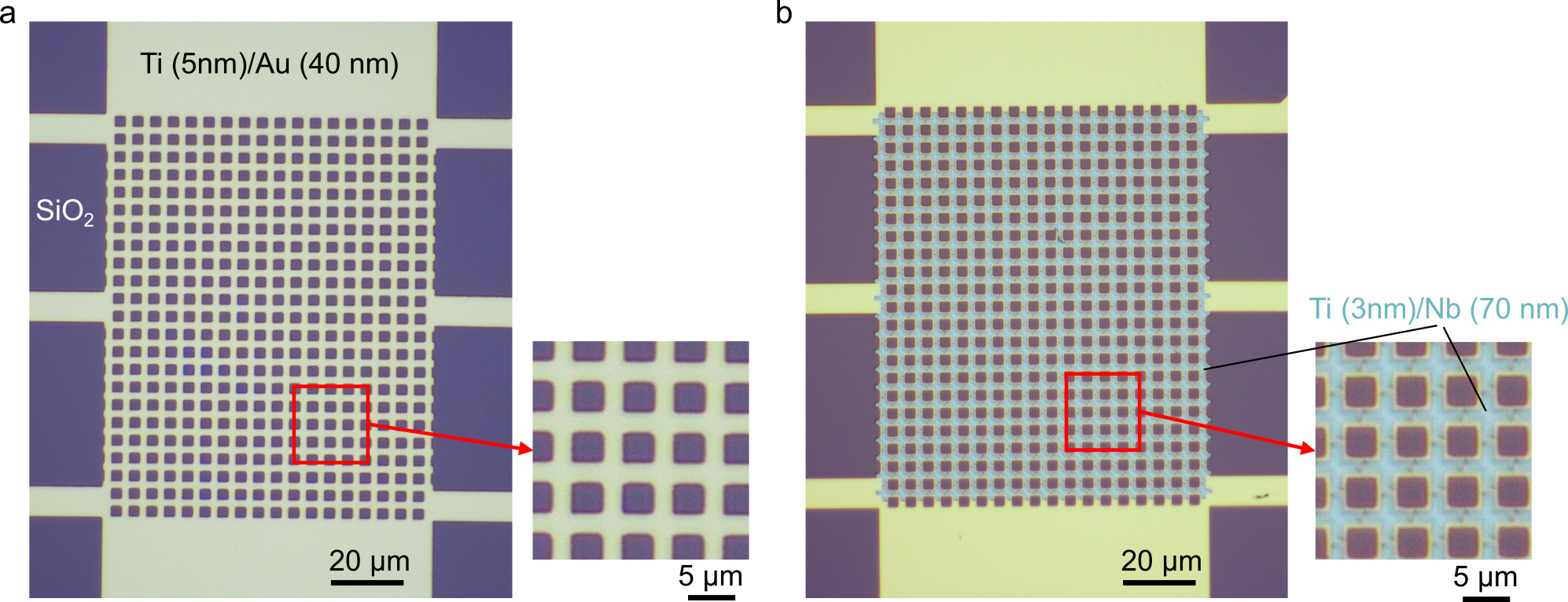}
\caption{\textbf{Fabrication of 2D Au-Nb Josephson junction arrays.} (a) Optical image of a Ti/Au Hall bar device with a continuous grid in the central region. (b) Optical image of a complete device. The blue crosses are the sputtered Ti/Nb islands.}
\label{fig:device}
\end{figure*}

\clearpage
\sisection{Fabrication of 2D Au-Nb Josephson junction arrays}
We describe the fabrication procedure for the Au-Nb Josephson junction arrays (JJAs). First, photolithography (Heidelberg MLA 150, AZ nLOF 2020 photoresist) was used to define the Hall-bar geometry and grid patterns on a Si/SiO$_2$ substrate, followed by electron-beam (e-beam) evaporation of Ti/Au (5/40 nm), as shown in Fig.~\ref{fig:device}(a). Next, e-beam lithography was performed to define the Nb island patterns. After resist development, a gentle oxygen-plasma cleaning step (O$_2$ flow rate: 42 sccm; power: 100 W; duration: 90 s) was applied to remove residual resist. The chip was then vacuum-sealed in a plastic bag using a vacuum sealer to minimize contamination before sputtering. Subsequently, Ti/Nb (3/70 nm) was deposited by DC magnetron sputtering \cite{Olszewski2026}. The thin Ti layer served to improve the adhesion of Nb to Au. Finally, lift-off was performed by sonicating the chip in acetone for 15 min and then in isopropyl alcohol (IPA) for another 15 min. An optical image of device S1 is shown in Fig.~\ref{fig:device}(b).

\sisection{Additional transport data}

\begin{figure*}[tbph]
\centering \includegraphics[width=0.8\textwidth]{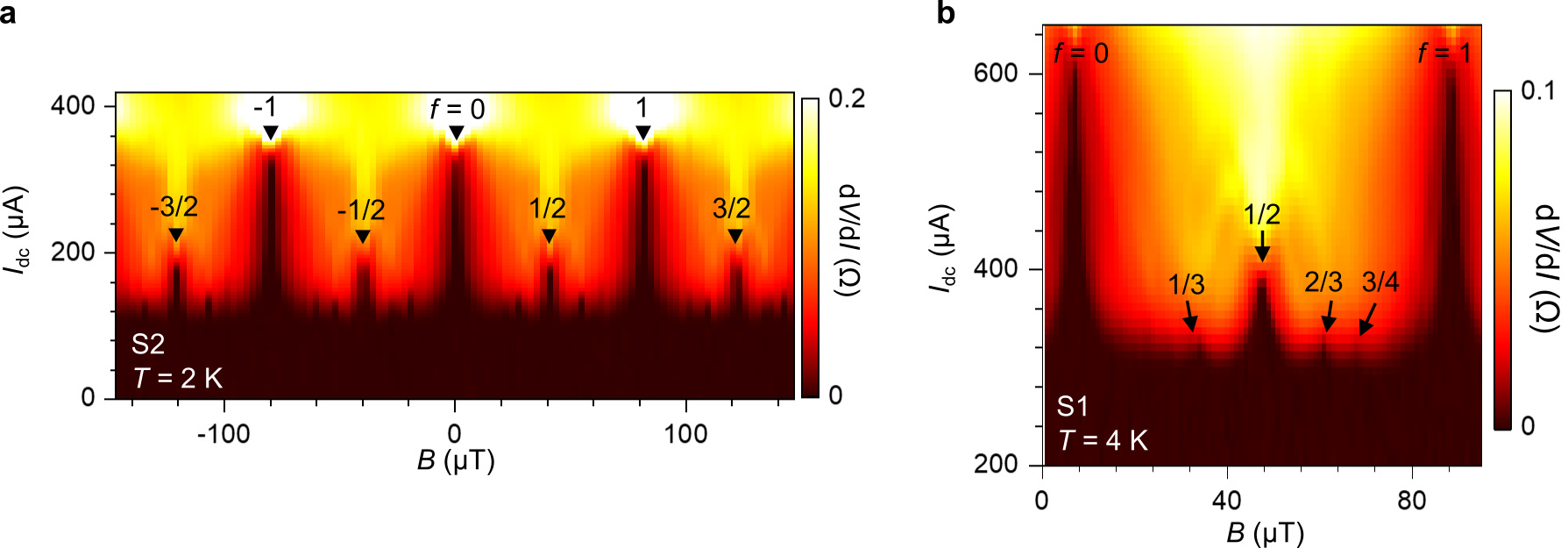}
\caption{\textbf{Transport characteristics of devices S2 and S1.} (a) Large-scale differential resistance $\mathrm dV/\mathrm dI$ map in device S2. (b) $\mathrm dV/\mathrm dI(B,I_{\mathrm{dc}})$ map for device S1 measured at \(T=4~\mathrm{K}\). The oscillation period is \(B_0\approx92~\mu\mathrm{T}\), in good agreement with \(B_0=\Phi_0/a^2=91.6~\mu\mathrm{T}\) for \(a=4.75~\mu\mathrm{m}\). Selected commensurate fillings are labeled.}
\label{fig:more_transport}
\end{figure*}

\begin{figure*}[tbph]
\centering \includegraphics[width=0.9\textwidth]{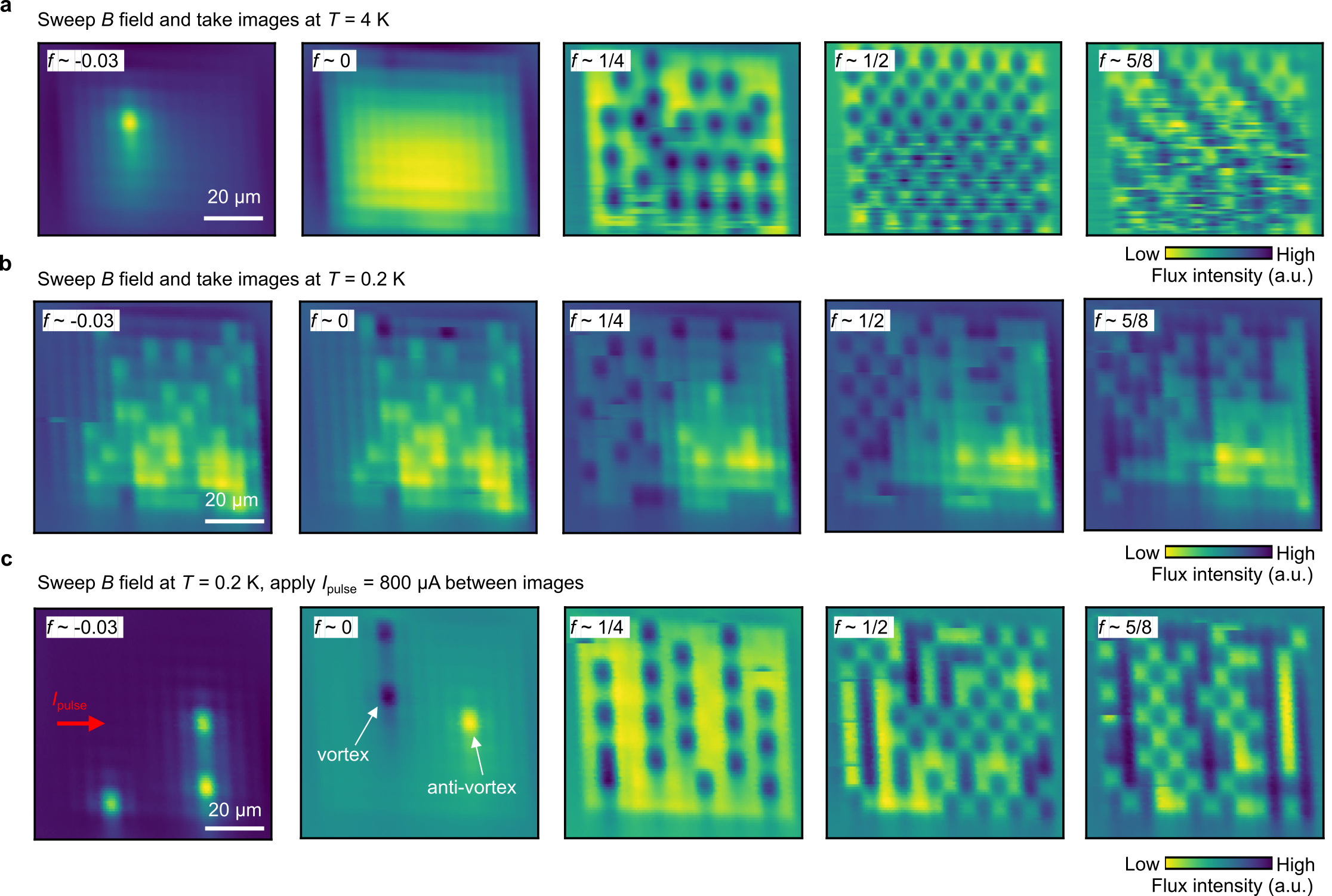}
\caption{\textbf{Comparison of measurement protocols for vortex imaging.} (a) Vortex images acquired after magnetic-field sweeps at $T=4~\mathrm{K}$. (b) Vortex images acquired after magnetic-field sweeps at $T=0.2~\mathrm{K}$. (c) Vortex images acquired at $T=0.2~\mathrm{K}$ after magnetic-field sweeps followed by a dc current pulse of amplitude $I_{\mathrm{pulse}}=800~\mu\mathrm{A}$ and duration $5~\mathrm{s}$. The red arrow indicates the direction of the applied current pulse.}
\label{fig:vortex-trapping}
\end{figure*}

\clearpage
\sisection{Temperature dependence of vortex mobility and imaging protocol}
We tested several protocols for imaging vortices in a representative JJA. First, we swept the magnetic field and imaged at $T=4~\mathrm{K}$, approximately $3~\mathrm{K}$ below $T_c^{\mathrm{arr}}$. At this temperature, vortices remain sufficiently mobile to form periodic configurations (see Fig.~\ref{fig:vortex-trapping}(a)). However, scanning the SQUID sensor perturbs the vortex positions, producing horizontal streaks in the image. We attribute this effect to local modification of the magnetic field by the superconducting shield of the SQUID chip.

We next cooled the array to $T=0.2~\mathrm{K}$, where the Josephson energy $E_J=\Phi_0 i_c/2\pi$ and the corresponding vortex pinning are substantially stronger. The scanning-induced streaks are strongly suppressed, but vortices become pinned and cannot readily reorganize following a magnetic-field sweep (Fig.~\ref{fig:vortex-trapping}(b)). Similar disordered configurations therefore persist over a range of fields, obscuring the commensurate states expected at $f=0$, $1/4$, $1/2$, and $5/8$.

We also tested current annealing at $T=0.2~\mathrm{K}$. Applying a dc current pulse above the array critical current drives vortices through the Lorentz force transverse to the current direction. Once the current is removed, however, the vortices become pinned again, resulting in the chain-like configurations shown in Fig.~\ref{fig:vortex-trapping}(c).

These observations demonstrate the strong temperature dependence of vortex mobility and motivate the field-cooling protocol used throughout the main text. At each magnetic field, we cool the array to $T<1~\mathrm{K}$ at a rate of $\sim1$ K/min before imaging. This allows vortices to reorganize while cooling, before pinning becomes strong, while ensuring that they are sufficiently immobilized during the subsequent SQUID scan to minimize sensor-induced rearrangement. This protocol reproducibly yields the commensurate vortex configurations presented in the main text.

\clearpage
\sisection{Simulations of the supercurrent and magnetic field distributions}

\begin{figure*}[tbph]
\centering \includegraphics[width=0.8\textwidth]{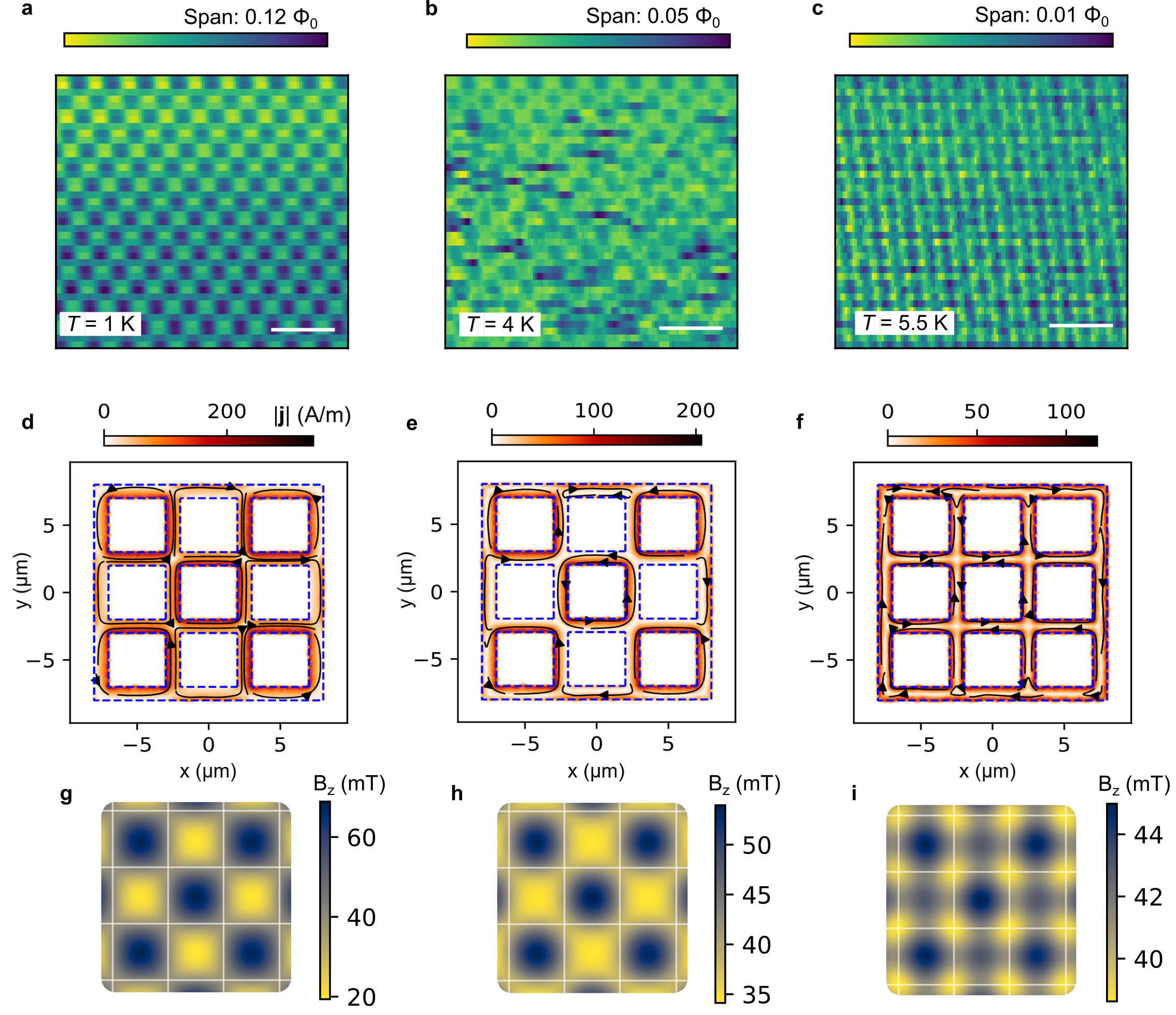}
\caption{\textbf{Supercurrent and magnetic-field distributions in a superconducting grid.} (a)–(c) Scanning SQUID flux images acquired in device S1 at $T=1$, 4, and 5.5 K, respectively. Scale bar, $20~\mu\mathrm{m}$. (d)–(f) Simulated supercurrent distributions in a $3\times3$ continuous superconducting grid at $f=1/2$, with alternating plaquette fluxes $r\Phi_0$ and $(1-r)\Phi_0$. We use (d) $r=0$, (e) $r=0.3$, and (f) $r=0.52$, estimated from the flux contrast in the SQUID images. A London penetration depth $\lambda=0.1~\mu\mathrm{m}$ is used; varying $\lambda$ does not change the qualitative behavior. (g)–(i) Corresponding magnetic-field distributions generated by the circulating currents in (d)–(f). A Gaussian filter is applied to account for the spatial resolution of the SQUID sensor. White lines indicate the superconducting grid. The magnetic-field maps are calculated over an extended $5\times5$ grid for clarity.}
\label{fig:screening}
\end{figure*}

In this section, we simulate the supercurrent distribution and the resulting magnetic-field profile in a continuous superconducting grid using the open-source Python package SuperScreen \cite{BishopVanHorn2022}. As shown in Figs.~\ref{fig:screening}(a)–(c), the contrast of the $f=1/2$ vortex pattern decreases strongly with increasing temperature, indicating that the flux contrast between neighboring plaquettes is reduced near the array transition temperature. To model this behavior, we consider a $3\times3$ square grid and assign alternating plaquette fluxes of $r\Phi_0$ and $(1-r)\Phi_0$, corresponding to a nominal filling of $f=1/2$. For $r=0$ (Fig.~\ref{fig:screening}(d)), neighboring plaquettes differ by one flux quantum and large circulating supercurrents develop around the cells. As $r$ approaches $1/2$ (Figs.~\ref{fig:screening}(e) and (f)), the flux contrast between neighboring plaquettes decreases, leading to a substantial reduction of the circulating current density and a more uniform current distribution along the superconducting grid. Figures~\ref{fig:screening}(g)–(i) show the corresponding magnetic-field profiles. As the plaquette-to-plaquette flux contrast decreases, the calculated magnetic-field modulation weakens, and the vortex features evolve from strongly lattice-confined shapes toward more rounded profiles. This evolution is consistent with the experimental observations. The accompanying reduction in circulating-current amplitude is also consistent with weaker vortex pinning as the Josephson coupling decreases at elevated temperature.

\clearpage
\sisection{Additional flux images at \texorpdfstring{$f=1/4$}{f=1/4}, \texorpdfstring{$2/5$}{2/5}, and \texorpdfstring{$3/7$}{3/7}}

\begin{figure*}[tbph]
\centering \includegraphics[width=0.9\textwidth]{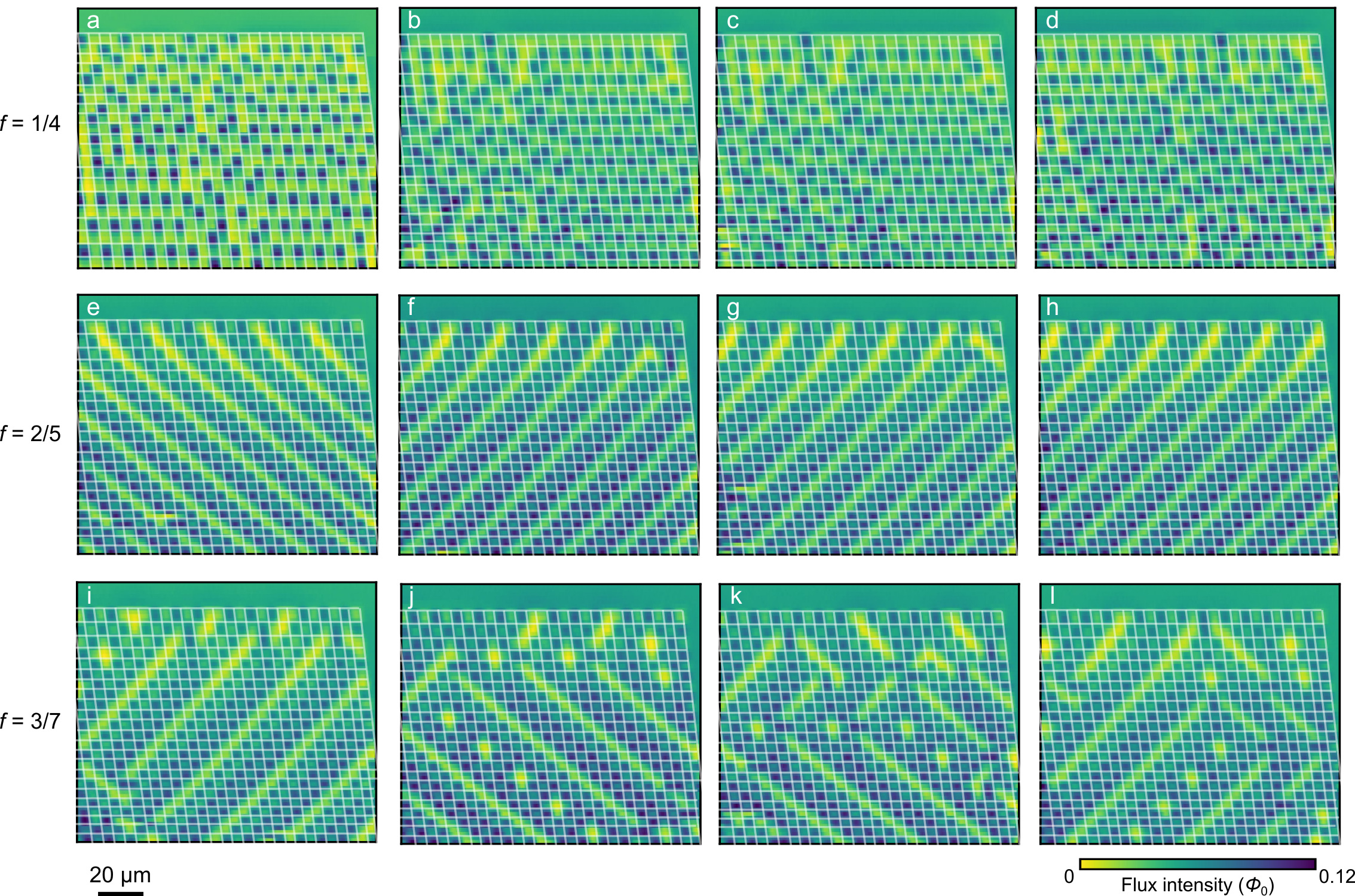}
\caption{\textbf{Additional vortex images at $f=1/4$, $2/5$, and $3/7$.} (a)–(d) Flux images at $f=1/4$, (e)–(h) at $f=2/5$, and (i)–(l) at $f=3/7$, acquired over multiple thermal cycles. White solid lines indicate the array lattice. All images are from device S1.}
\label{fig:more_images}
\end{figure*}

Figure~\ref{fig:more_images} shows additional flux images acquired over repeated thermal cycles at $f=1/4$, $2/5$, and $3/7$. The striped $f=2/5$ state is highly reproducible (Figs.~\ref{fig:more_images}(e)–(h)), with domains appearing in symmetry-related orientations connected by the $C_4$ symmetry of the square array Hamiltonian \cite{teitel1983josephson,martinoli2000two}. At $f=3/7$, we occasionally observe striped domains that can be viewed as vacancy superlattices, separated by domain walls running along the array diagonals.

\begin{figure*}[tbph]
\centering \includegraphics[width=0.9\textwidth]{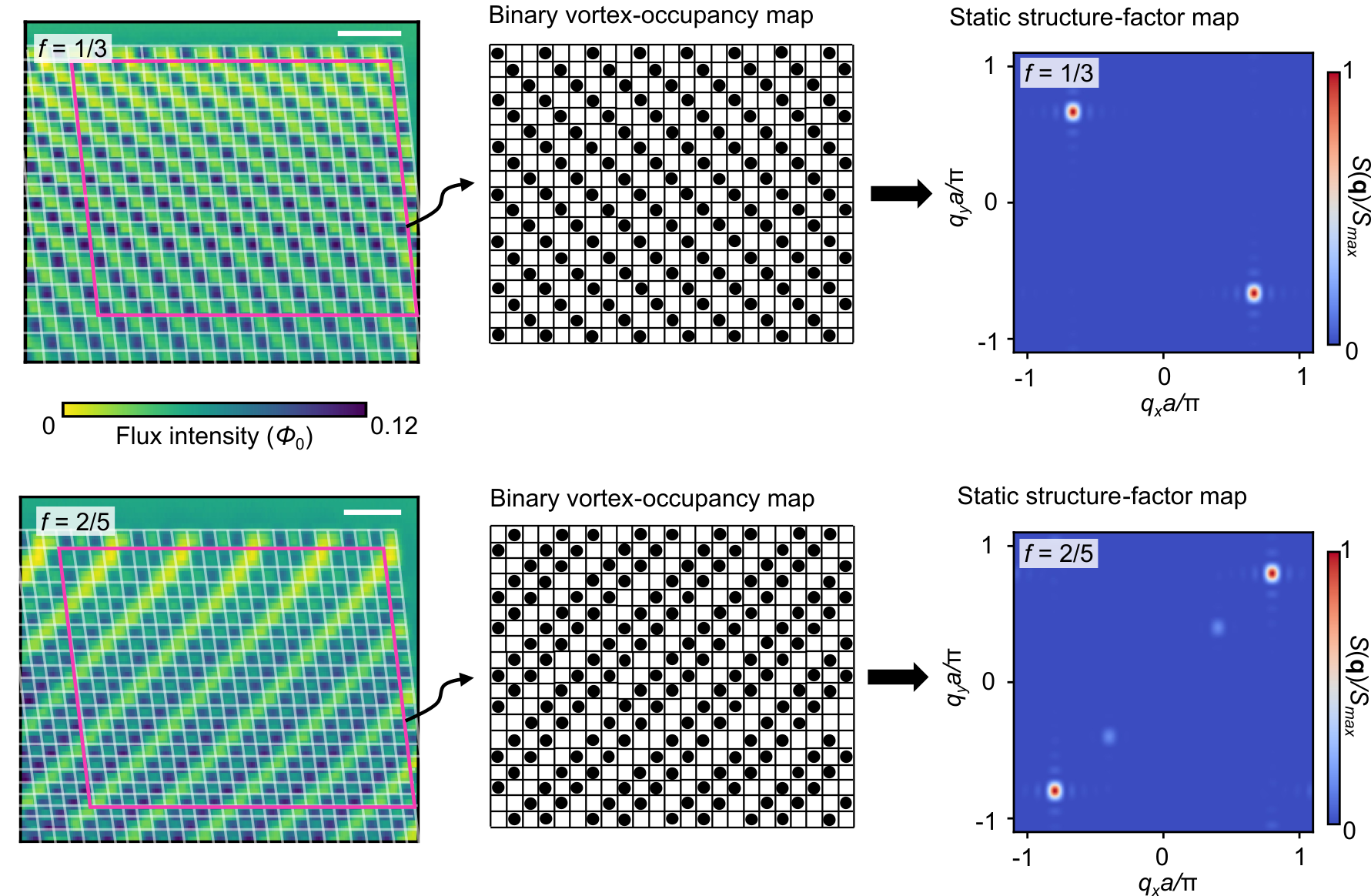}
\caption{\textbf{Structure-factor analysis of additional $f=1/3$ and $2/5$ states.}
Left panels: real-space flux images acquired at $f=1/3$ and $2/5$. Middle panels: extracted binary vortex-occupancy maps from the fixed $N_x=19$ by $N_y=23$ plaquette windows indicated by the magenta boxes. Right panels: corresponding static structure-factor maps $S(\mathbf q)$ within the first Brillouin zone.}
\label{fig:SF_1}
\end{figure*}

\clearpage
\sisection{Structure-factor calculation and analysis}
To characterize the spatial ordering of the imaged vortex configurations, we calculate the static structure factor $S(\mathbf q)$ from binary vortex-occupancy maps extracted from the scanning SQUID images. For all datasets, we analyze the same finite window containing $N_x=19$ by $N_y=23$ plaquettes (total number of plaquettes $N=437$), as illustrated in Fig.~\ref{fig:SF_1}. Each plaquette is assigned an occupation number $n_i=1$ if it contains a vortex and $n_i=0$ otherwise. We define
\begin{equation}
S(\mathbf q)=\frac{1}{N}\left|\sum_{i=1}^{N}(n_i-\bar n)e^{i\mathbf q\cdot\mathbf r_i}\right|^2,
\end{equation}
where $\bar n=N^{-1}\sum_i n_i$ is the mean vortex density, and $\mathbf r_i$ is the position of the $i$-th plaquette. For the square arrays, the plaquette centers are taken as $\mathbf r_i=a(x_i,y_i)$, where $a$ is the array lattice constant and $x_i$ and $y_i$ are integers. We numerically evaluate the discrete Fourier transform of $n_i-\bar n$ over a dense reciprocal-space grid spanning the first Brillouin zone, $-\pi/a\leq q_x,q_y\leq\pi/a$. Because the analysis is performed over a finite field of view, the Bragg peaks have a finite reciprocal-space width even for an ideal configuration. Figure~\ref{fig:SF_1} shows representative structure-factor maps at $f=1/3$ and $2/5$.

\begin{figure*}[tbph]
\centering \includegraphics[width=0.55\textwidth]{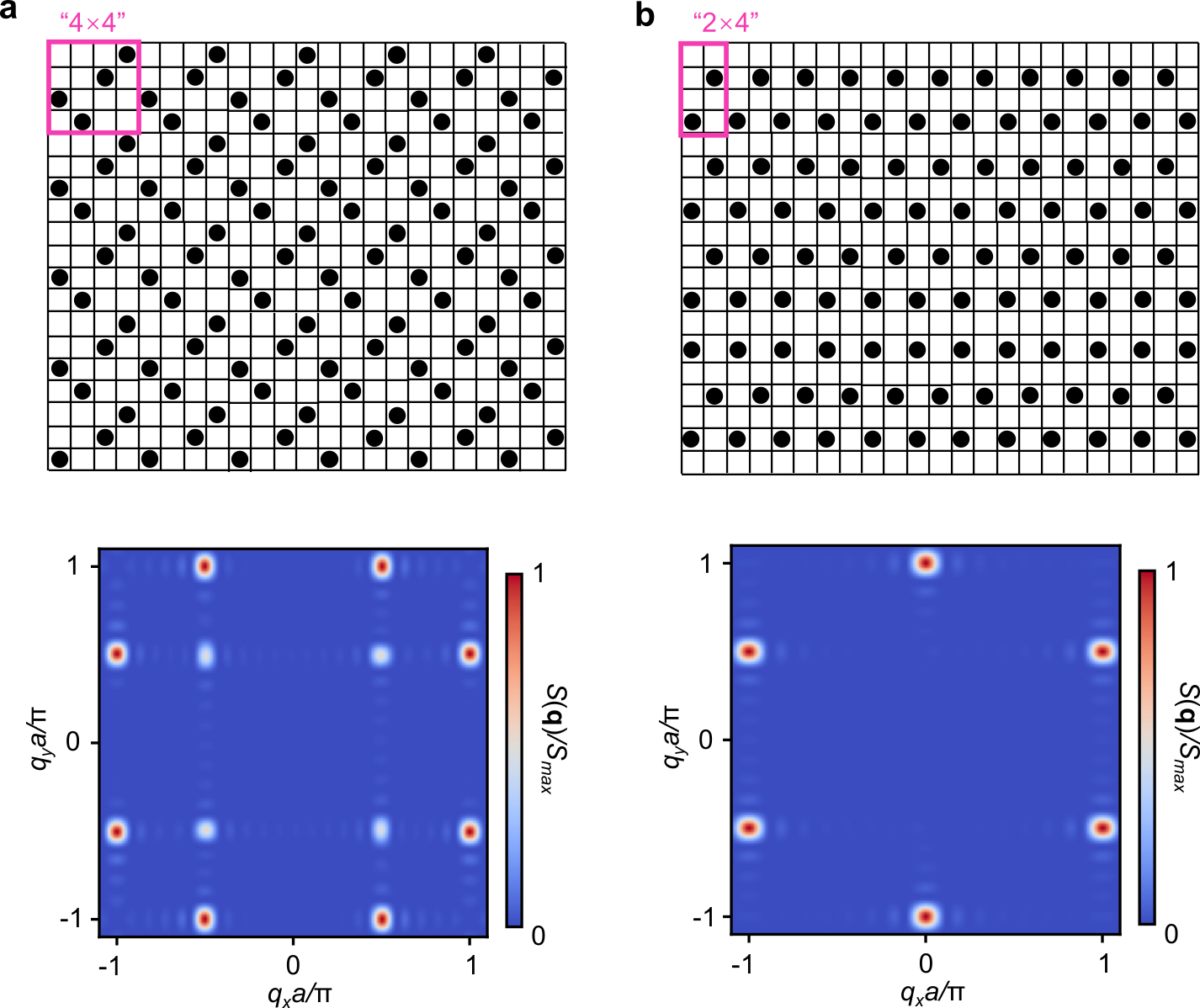}
\caption{\textbf{Structure-factor signatures of the competing $f=1/4$ vortex configurations.} (a) Ideal $4\times4$ vortex configuration constructed over the same $N_x=19$ by $N_y=23$ plaquette window used for the experimental analysis, together with its calculated structure-factor map $S(\mathbf q)$ within the first Brillouin zone. (b) Corresponding analysis for the ideal $2\times4$ vortex configuration.}
\label{fig:SF_2}
\end{figure*}

To identify the competing orders at $f=1/4$, we calculate the structure factors of ideal $4\times4$ and $2\times4$ vortex configurations (Fig.~\ref{fig:SF_2}). The $4\times4$ configuration produces strong Bragg peaks at $(q_xa/\pi,q_ya/\pi)=(\pm1,\pm1/2),\;(\pm1/2,\pm1)$, together with secondary peaks at $(\pm1/2,\pm1/2)$. The $2\times4$ configuration has characteristic peaks at $(q_xa/\pi,q_ya/\pi)=(0,\pm1),\;(\pm1,\pm1/2)$. The experimental structure-factor map at \(f=1/4\), shown in Fig. 2(h) of the main text, contains features from both sets of ordering wavevectors, providing reciprocal-space evidence for coexistence of the two configurations. Its broader Bragg peaks and stronger diffuse scattering indicate reduced spatial coherence associated with competing domains.

\begin{figure*}[tbph]
\centering \includegraphics[width=0.9\textwidth]{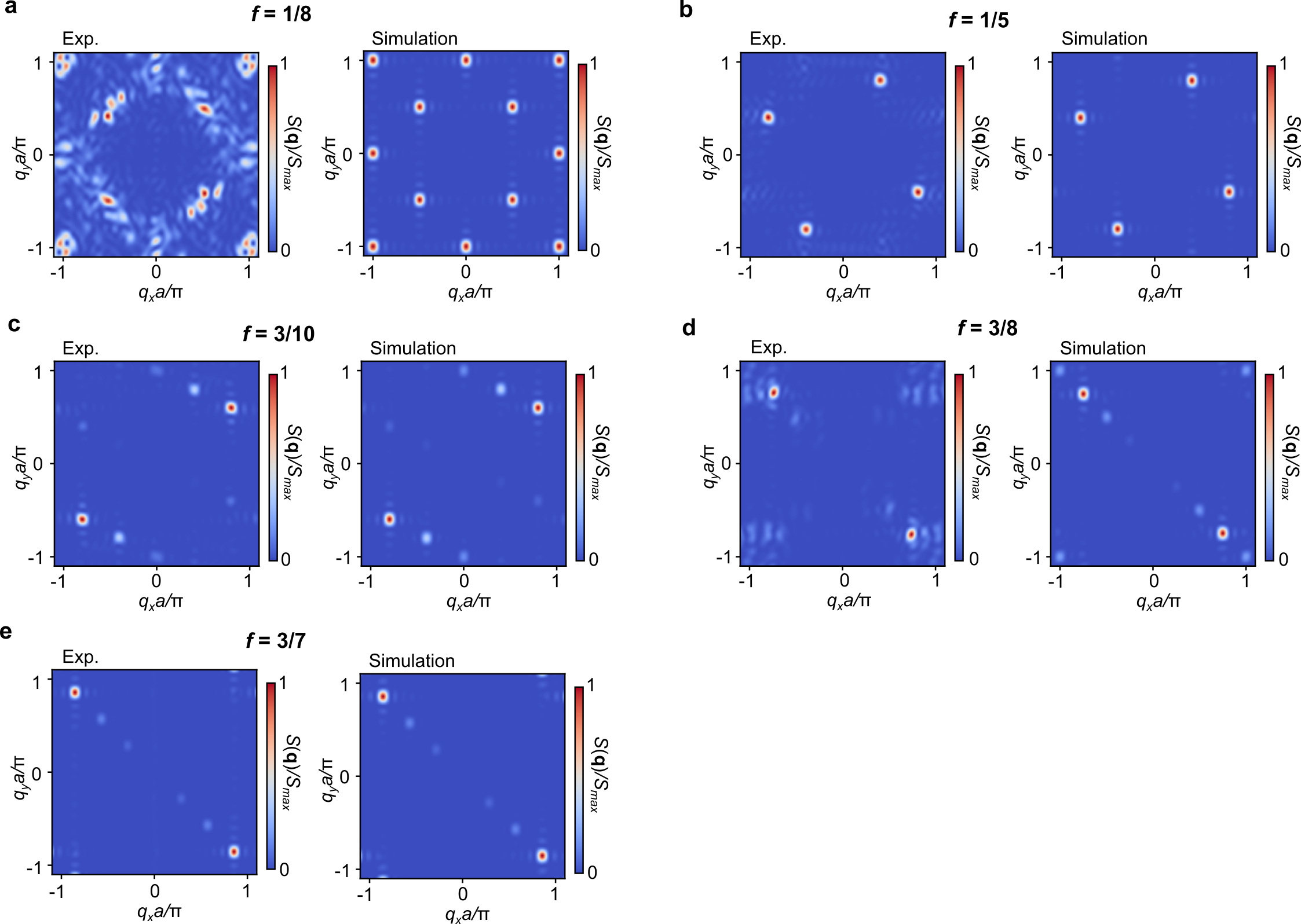}
\caption{\textbf{Comparison of experimental and ideal structure factors for commensurate vortex states.}}
\label{fig:SF_3}
\end{figure*}

Figure~\ref{fig:SF_3} compares the experimental structure factors corresponding to Fig. 3(a)--(e) of the main text with those calculated for ideal vortex lattices at $f=1/8$, $1/5$, $3/10$, $3/8$, and $3/7$. The experimental maps retain the characteristic Bragg-peak patterns of the corresponding ideal configurations, supporting the assignment of the observed vortex states. Additional peaks, broadening, and diffuse reciprocal-space weight reflect disorder, finite domain size, and reduced long-range coherence.

To quantify the stability of commensurate order away from the optimal filling, we use the normalized principal Bragg-peak intensity
\begin{equation}
\eta(f) = \frac{S_f(\mathbf Q)} {S_{\mathrm{ideal}}(\mathbf Q)},
\end{equation}
where \(\mathbf Q\) is the principal ordering wavevector and $S_{\mathrm{ideal}}(\mathbf Q)$ is calculated for an ideal, defect-free vortex lattice over the same finite field of view. This normalization gives $\eta=1$ for the corresponding ideal configuration. For the $f=1/2$ checkerboard state, $\mathbf Q=(\pi/a,~\pi/a)$. For $f=1/3$, two symmetry-related stripe orientations are possible, and we characterize the dominant long-range order using
\begin{equation}
\eta_{1/3}(f) = \frac{ \max[ S_f(\mathbf Q_1), S_f(\mathbf Q_2) ] }{ \max[ S_{\mathrm{ideal}}(\mathbf Q_1), S_{\mathrm{ideal}}(\mathbf Q_2) ] },
\end{equation}
where $\mathbf Q_1=(2\pi/3a,~2\pi/3a)$ and $\mathbf Q_2=(2\pi/3a,~-2\pi/3a)$
are the principal ordering wavevectors of the two stripe orientations, related by a $90^\circ$ rotation.

\sisection{Local analysis of competing vortex orders at \texorpdfstring{$f=1/4$}{f=1/4}}
To quantify the local coexistence of the $4\times4$ and $2\times4$ vortex configurations at $f=1/4$, we perform template matching on the binary vortex-occupancy maps using overlapping $4\times4$-plaquette windows. Ideal $4\times4$ and $2\times4$ patterns, including symmetry-equivalent translations and $90^\circ$ rotations, are compared with the experimental configuration within a sliding local window. At each position, we calculate the normalized correlation
\begin{equation}
W= \frac{\sum_j (n_j-\bar n)(t_j-\bar t)} {\sqrt{\sum_j(n_j-\bar n)^2} \sqrt{\sum_j(t_j-\bar t)^2}},
\end{equation}
where $n_j$ and $t_j$ are the vortex occupancies in the experimental window and ideal template, respectively, and $\bar n$ and $\bar t$ are their mean occupancies. We define $W_{4\times4}$ and $W_{2\times4}$ as the maximum correlations over all symmetry-equivalent templates. A window is assigned to the $4\times4$ ($2\times4$) configuration only if $W_{4\times4}\ge 0.9$ ($W_{2\times4}\ge 0.9$) and the corresponding correlation score exceeds that of the other configuration by at least 0.1. Windows that do not satisfy both criteria are classified as disordered or ambiguous.

The resulting classification yields the fractions $F_{4\times4}=N_{4\times4}/N_{\mathrm{tot}}$ and $F_{2\times4}=N_{2\times4}/N_{\mathrm{tot}}$, with $F_{\mathrm{other}}=1-F_{4\times4}-F_{2\times4}$ accounting for domain boundaries, defects, and ambiguous regions. Applying this analysis to the SQUID images shown in Fig. 2(d) of the main text and Figs.~\ref{fig:more_images}(a)--(d), we obtain average $F_{2\times4}\approx33\pm4\%$, $F_{4\times4}\approx26\pm3\%$, and $F_{\mathrm{other}}\approx41\pm2\%$. The reported fractions are not sensitive to the specific classification criteria: varying the minimum-correlation threshold and the required difference between the two correlation scores over a reasonable range produces only minor changes in the extracted fractions.

\clearpage
\sisection{Monte Carlo simulations of commensurate vortex configurations}
\label{sec:xy_simulations}
\sisubsection{Model and numerical procedure}
We describe the square JJA using the classical uniformly frustrated XY model~\cite{teitel1983josephson,halsey1985josephson}, with a superconducting phase $\theta_i$ on each island and nearest-neighbor Josephson coupling:
\begin{equation}
H=-\sum_{\langle ij\rangle}E_{J,ij}\cos\left(\theta_i-\theta_j-A_{ij}\right),
\qquad \sum_{\partial p}A_{ij}=2\pi f\pmod{2\pi}.
\label{eq:xy_hamiltonian}
\end{equation}
Each bond is counted once, $A_{ji}=-A_{ij}$, and the oriented sum is around a plaquette $p$. We use the Landau gauge, $A_x(x,y)=0$ and $A_y(x,y)=2\pi fx$, with integer site coordinates, and set $E_{J,ij}=E_J$ (uniform and isotropic coupling) unless otherwise specified. The calculation neglects magnetic self-screening, charging effects, and applied transport current. Periodic boundary conditions represent the interior of an ideal array; for $f=p/q$, we choose $L$ to be a multiple of $q$ on an $L\times L$ lattice. One phase is fixed to remove the freedom of a uniform phase rotation, i.e., the global $U(1)$ freedom. Energies per site are $\varepsilon=H/N$, where $N=L^2$.

Starting from independent phases uniformly distributed over $[0,2\pi)$, we perform single-site Metropolis updates. A trial change $\theta_i\rightarrow\theta_i+\delta\theta$ takes the phase configuration $\boldsymbol{\theta}$ to $\boldsymbol{\theta}'$ and is accepted with probability
\begin{equation}
P_{\mathrm{acc}}=\min\!\left[1,\exp\!\left(-\frac{H(\boldsymbol{\theta}')-H(\boldsymbol{\theta})}{k_BT}\right)\right].
\label{eq:xy_metropolis}
\end{equation}
Here, $H(\boldsymbol{\theta}')$ and $H(\boldsymbol{\theta})$ are the energies of the trial and current configurations evaluated using the same Hamiltonian, Eq.~\eqref{eq:xy_hamiltonian}. One sweep visits every unfixed site once in random order. Trial phase changes are drawn uniformly from $[-\delta\theta_{\max},\delta\theta_{\max}]$, with the finite-temperature proposal width chosen as $\delta\theta_{\max}=\min[\pi,3\sqrt{k_BT/E_J}]$ to reduce the step size during cooling, chosen for numerical stability. For the commensurate states discussed below, we cool geometrically from $k_BT/E_J=2$ to $10^{-4}$ over $2\times10^4$ sweeps, using six independent random seeds and sizes $L=2q$ and $4q$. For $f=2/5$ at $L=20$, we additionally use $2\times10^5$ sweeps to reduce trapping in configurations containing defects. After annealing to the finite temperature specified above, we perform a separate zero-temperature energy minimization with analytic phase gradients; this step does not use Monte Carlo proposals. The maximum residual $|\partial H/\partial\theta_i|$ is below $10^{-6}E_J$. We compare the lowest energies reached across runs, since an individual anneal does not necessarily reach the lowest-energy state.

Vortex occupations are extracted from the wrapped gauge-invariant phase differences,
\begin{equation}
n_p=f+\frac{1}{2\pi}\sum_{(ij)\in\partial p}
\operatorname{wrap}_{[-\pi,\pi)}\!\left(\theta_i-\theta_j-A_{ij}\right).
\label{eq:xy_vorticity}
\end{equation}
The result is integer-valued up to numerical precision. The ordered states considered here have $n_p=0$ or $1$. These maps allow direct comparison of the vortex arrangements with the experimentally identified unit cells.

\sisubsection{Commensurate states at $f=1/2$, $1/3$, and $2/5$}
The lowest-energy configurations recovered at both tested sizes reproduce the checkerboard at $f=1/2$ and the diagonal staircase arrangements at $f=1/3$ and $2/5$. The latter consist of one occupied diagonal in a three-diagonal period and two occupied diagonals in a five-diagonal period, respectively. Translations and symmetry-related stripe orientations yield equivalent patterns. At both sizes, the relaxed energies per site agree to numerical precision with $-\sqrt{2}E_J$, $-\frac{4}{3}E_J$, and $-\frac{2(1+\sqrt{5})}{5}E_J$, respectively. The resulting unit cells agree with those identified in Fig.~2(e)--(g) of the main text, supporting the interpretation of the observed patterns as commensurate vortex order in the frustrated XY model.

\sisubsection{Competing configurations at $f=1/4$}
To compare the two motifs in Fig.~\ref{fig:SF_2}, we initialize phases corresponding to each ideal vortex pattern, add small independent Gaussian phase perturbations of standard deviation $0.05$ radians, and relax on $L=8$ and $16$ lattices. Initial phases are obtained from a discrete Poisson solve for a bond-phase field with the prescribed plaquette circulation; subsequent updates and minimization use the full cosine Hamiltonian. For each size and motif, three independent seeds are subjected to 100 Metropolis sweeps from $k_BT/E_J=0.02$ to $10^{-5}$, followed by the same zero-temperature minimization. No vortex-position constraints are imposed during relaxation. Both motifs are preserved and give
\begin{equation}
\varepsilon_{4\times4}=\varepsilon_{2\times4}
=-\frac{1+\sqrt{3}}{2}E_J.
\label{eq:xy_quarter_energy}
\end{equation}
The phase Hessian has only the uniform-phase zero mode and positive remaining eigenvalues for both configurations, establishing their local stability. Their equal energies are consistent with the coexistence and comparable ordered-area fractions observed experimentally.

The distinction between vortex periodicity and phase periodicity is essential in this comparison. Although the $2\times4$ vortex pattern fits within a $4\times4$ cell, its optimal phases require a $\pi$ twist across that cell in one direction. Ordinary periodic phase boundaries at $L=4$ therefore bias the comparison. In contrast, both optimal configurations are accommodated at $L=8$ and $16$. The equality in Eq.~\eqref{eq:xy_quarter_energy} refers to the zero-temperature energies. For completeness, we compare Gaussian phase fluctuations about the two minima using
\begin{equation}
\Delta F_{\mathrm{harm}}=F_{2\times4}-F_{4\times4}
=\frac{k_BT}{2}\sum_{\alpha}'\ln\!\left[\frac{\lambda_{\alpha}^{(2\times4)}}{\lambda_{\alpha}^{(4\times4)}}\right],
\label{eq:xy_harmonic}
\end{equation}
where $\lambda_\alpha$ are the eigenvalues of the phase Hessian. The prime restricts the sum to the $N-1$ nonzero modes, excluding the zero mode associated with a uniform rotation of all phases. This low-temperature approximation gives a weak free-energy preference for the $4\times4$ motif.

We also test weak directional anisotropy by replacing the uniform coupling with $E_{J,x}=E_J(1+\delta)$ and $E_{J,y}=E_J(1-\delta)$, where $|\delta|\ll1$, and independently minimizing the $4\times4$ state and both orientations of the $2\times4$ state. For either sign of the anisotropy, one orientation of the $2\times4$ state is favored, with the preferred orientation interchanged when the sign of $\delta$ is reversed. This could be a mechanism for the slightly larger observed $2\times4$ area fraction.

\end{document}